\documentclass[twocolumn,showpacs,floatfix,prb,10pt]{revtex4-1}
\usepackage[english]{babel}
\usepackage{amsmath,amssymb}
\usepackage{times}
\usepackage{epsfig}
\usepackage[colorlinks,linkcolor=blue,citecolor=blue]{hyperref}
\usepackage{float}
\begin{document}
\title{Light induced magnetization in a spin $S=1$ easy--plane antiferromagnetic chain}
\author{J. Herbrych$^{1}$}
\author{X. Zotos$^{1,2,3}$}
\affiliation{$^1$Cretan Center for Quantum Complexity and Nanotechnology and 
Institute of Theoretical and Computational Physics, Department of Physics, 
University of Crete, Heraklion 71003, Greece}
\affiliation{$^2$Foundation for Research and Technology - Hellas, 71110 Heraklion, Greece}
\affiliation{$^3$ Max-Planck-Institut f\"ur Physik komplexer Systeme, 
N\"othnitzer Strasse 38, 01187 Dresden, Germany}
\date{\today}
\begin{abstract}
The time evolution of magnetization induced by circularly polarized light in a $S=1$ Heisenberg 
chain with large easy--plane anisotropy is studied numerically and analytically. Results at constant 
light frequency $\Omega=\Omega_0$ are interpreted in terms of absorption lines of the electronic 
spin resonance spectrum. The application of time dependent frequency $\Omega=\Omega(t)$ light, so 
called chirping, is shown to be an efficient procedure in order to obtain within a short time a 
large, controlled value of the magnetization $M^z$. Furthermore, comparison with a $2$--level model 
provides a qualitative understanding of the induced magnetization process.
\end{abstract}
\maketitle

Far-from-equilibrium condensed matter physics is a challenging, still largely uncharted territory. 
Concerning the out of equilibrium dynamics of quantum magnets, the control of magnetic properties by 
means other than a conventional magnetic field is of strong current interest
\cite{Asamitsu1997,Lottermoser2004,Kato2004,Kimel2005,Hild2014}. 
For instance, engineering the quantum state, i.e. wavefunction, is essential for quantum simulators, 
precision sensors or spintronic devices \cite{Cole2001,Nayak2008,Simon2011,Britton2012,Stern2013}. 
Recent experimental advances allow to manipulate the elementary low--energy excitations with terahertz 
laser pulses \cite{Huber2001,Beaurepaire2004,Takahashi2009,Kirilyuk2010,Kampfrath2011}, a prominent 
example being the ultrafast coherent control of antiferromagnetic magnons. A time--dependent (
rotating) magnetic field of highly intense terahertz laser pules, with photon energy below the 
electron energy scale, controlled the coherent spin waves without interfering with the motion of 
charge carriers.

In quantum magnets with reduced dimensionality the thermodynamic and transport properties exhibit a 
rich magnetic field dependence
\cite{Sologubenko2007a,Sologubenko2007b,Sologubenko2008,Sologubenko2009, Zapf2006,Klanjsek2008, Sun2009,Kohama2011} 
related to the total magnetization of the system. Prominent examples of such a behaviour are the 
field--induced quantum phase transitions of the organic compound \mbox{NiCl$_2$-4SC(NH$_2$)$_2$} 
(dichlorotetrakisthiourea--nickel abbreviated as DTN). At zero temperature, the first transition 
occurs at a critical field $h_1$ where the energy gap closes and a finite magnetization develops in 
the ground state (GS); the second one occurs at $h_2$ where the magnetization fully saturates 
leading to a ferromagnetic GS. By now, the low--energy physics of the DTN compound has been well studied 
experimentally \cite{Zapf2006,Zvyagin2007,Zvyagin2008,Sun2009,Kohama2011,Mukhopadhyay2012} and 
understood theoretically. The basic model that describes the magnetic excitation spectrum of DTN was 
found to be the one dimensional $S=1$ antiferromagnetic Heisenberg model (AHM) with exchange 
coupling constant $J$ and large easy--plane anisotropy $D$. As shown in 
Refs.~\onlinecite{Papanicolaou1997,Psaroudaki2012,Psaroudaki2013,Zapf2014} such a Hamiltonian 
reproduces in great detail the low lying electronic spin resonance (ESR) spectrum. The anisotropy 
$D/J \sim 4$ of DTN, being the largest energy scale in the system, is responsible for a large energy 
gap ${\cal O}(D)$ that can be closed by a magnetic field $h$. 

In this work we study the rotating magnetic field induced nonequilibrium magnetization $M^z$ in 
large, easy - plane anisotropy AHM. For a field rotating at constant frequency (circularly polarized 
light) we are connecting the numerical results with the linear response (LR) theory predictions for 
the transition frequency of the corresponding ESR experiment. In the case of a chirped (time dependent) 
frequency of the light \cite{Zamith2001}, our results indicate that the short--time behaviour of the 
magnetization is mainly driven by the anisotropy part of the system. This time scale, together with 
the dependence of the magnetization on the chirp parameters, can be accurately described by a $2$--
level model. The dynamics beyond the characteristic time of the latter is dominated by the 
Heisenberg part of the model. Although we focus on DTN as a typical one dimensional $S=1$ easy--
plane AHM, our analysis is also valid for other Hamiltonians, e.g. a $2$--level model will yield 
the correct physics for $S=1$ models in all dimensions provided that $D\gg J$. 

As prototype model we choose the $S=1$ AHM with single--site, easy--plane anisotropy $D$ on a chain 
with $L$ sites
\begin{equation}
H_0=\sum_{i=1}^{L}\left[J\mathbf{S}_i\cdot\mathbf{S}_{i+1}+D(S^z_i)^2+hS^z_i\right]\,,
\label{s10} 
\end{equation} where $\mathbf{S}_i=(S^x_i,S^y_i,S^z_i)$ are spin $S=1$ operators at site $i$,
$\mathbf{S}_{L+1}=\mathbf{S}_{1}$ (periodic boundary conditions), $h$ is a magnetic field, and
$J (\sim 2\,\text{K})$ the antiferromagnetic exchange constant (we will further on use
$\hbar=k_B=\mu_B=1$ and set $J=1$ as the unit of energy). 
Hereafter, we will use $D=4~(\sim 8\text{K})$ and for such 
an anisotropy the critical fields are: $h_1\simeq 2.28$ and $h_2=8$ \cite{Psaroudaki2012}. 
We will assume that only the 
magnetic component of light, propagating in the $z$--direction, couples to the system. The time--
dependent Hamiltonian of the corresponding setup can be written as
\begin{equation}
H(t)=H_0-A\sum_{i=1}^{L}\left(\mathrm{e}^{-\dot{\iota}\Omega t}S^{+}_i+ \mathrm{e}^{\dot{\iota}\Omega t}S^{-}_i\right)\,,
\label{s1t}
\end{equation}
where $A>0$ and $\Omega>0$ are the amplitude and frequency of light respectively and $S^{\pm}_{i}$ 
are spin raising and lowering operators. Thus, each spin ``feels'' a magnetic field rotating in the
$xy$--plane, $2A\sum_{i}[S^{x}_{i}\cos(\Omega t)+S^{y}_{i}\sin(\Omega t)]$. The magnetization 
induced is positive, in order to obtain a negative magnetization one should 
substitute $\Omega\to-\Omega$ in \eqref{s1t}. Note that in a real experiment a propagating light 
pulse has some time and frequency dependence, an issue that we will discuss later on. In order to 
probe the sample magnetization perpendicular to the polarization plane one can use a second optical 
pulse and measure the change in its polarization state induced by the magnetization either in 
transmission (Faraday effect) or reflection (Kerr effect) geometry.
\cite{Kato2004,Ziel1965,Kampfrath2011}

The time evolution of the magnetization is given by
\begin{equation}
M^z(t)=\frac{\langle\Psi(t)|S^z_{\text{tot}}|\Psi(t)\rangle}{\langle\Psi(t)|\Psi(t)\rangle}\,,
\label{magev}
\end{equation}
where $S^z_{\text{tot}}=(1/L)\sum_i S^z_i$ and $|\Psi(t)\rangle$ is a solution of the
time--dependent Schr\"{o}dinger equation
$\dot{\iota} \partial_t|\Psi(t)\rangle=H(t)|\Psi(t)\rangle$. In our calculation we choose $\delta t$ 
in such a way that typically $\langle \Psi(t)|\Psi(t)\rangle\simeq 1$ at any time $t$ 
($\delta t\simeq 10^{-3}$). A general procedure goes as follows: (i) first, with help 
of exact diagonalization we calculate the GS of \eqref{s10}, 
$|\Psi(-\delta t)\rangle =|\text{GS}\rangle$. (ii) Next, at time $t=0$ we instantaneously turn on 
the light and (iii) finally, we perform the time evolution of it on the basis of the 
time--discretized version of the Schr\"{o}dinger equation with \eqref{s1t} (using a fourth--order 
Runge--Kutta routine).

Let us first focus on the system \eqref{s1t} at constant frequency $\Omega=\Omega_0$. It is clear 
that the maximum value of the magnetization is induced by light at the resonance frequency of 
the system, $\Omega_0=\Omega_R$, that can be interpreted in the spirit of an ESR spectrum. For 
small enough $A~(\ll J)$ the system is in the linear response regime and the low absorption 
lines of the ESR spectrum of \eqref{s10} \cite{Oshikawa1999,Psaroudaki2012} correspond to the 
resonance frequencies $\Omega_R$ of \eqref{s1t} at given $h$. Furthermore, \eqref{s1t} at 
$\Omega=\Omega_0$ can be mapped by a unitary transformation (or Floquet theory) to an effective 
static model \cite{Takayoshi1,Takayoshi2,3435}, where the latter has a form similar to the one when 
dealing with an ESR experiment. Note that the same procedure was used in Ref.~\onlinecite{Takayoshi1} 
in order to study a system with small magnetic anisotropy $D=0.25$ (Haldane-like limit).

\begin{figure}[!ht]
\includegraphics[width=0.9\columnwidth]{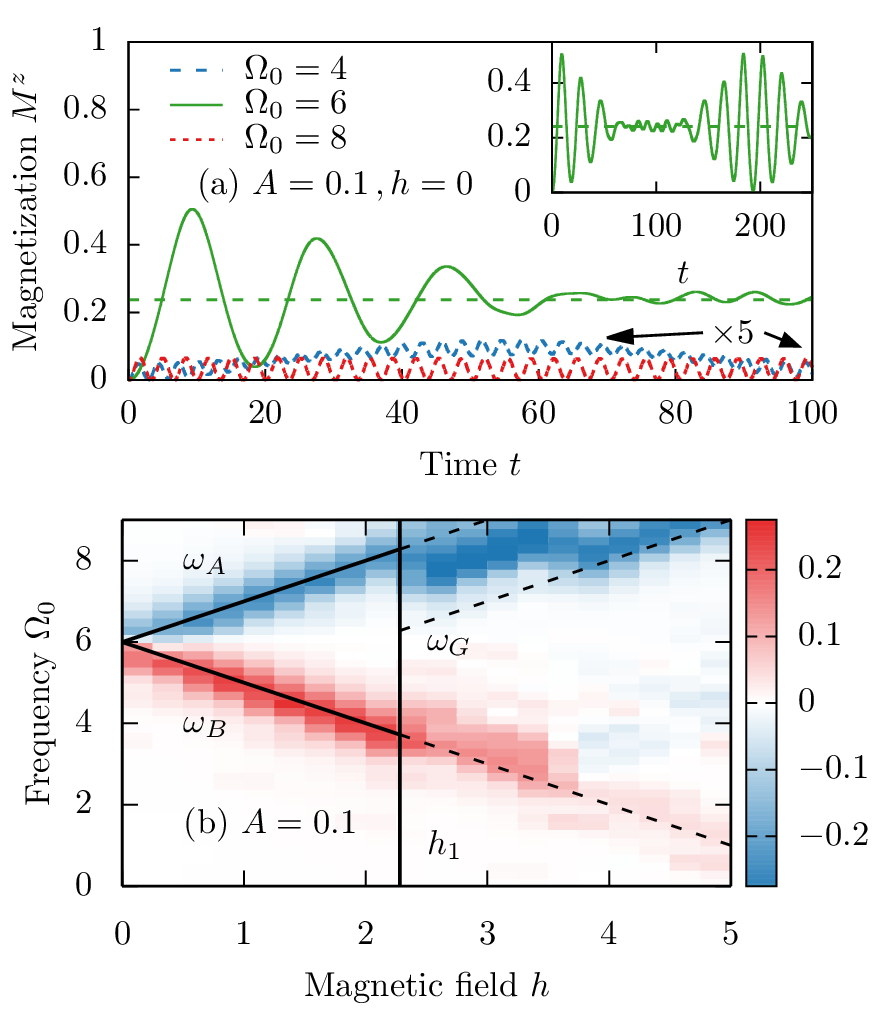}
\caption{(Color online) (a) Magnetization as a function of time $M^z(t)$ calculated for $L=11$, 
$A=0.1$, $h=0$, and $\Omega_0=4,6,8$. Dashed horizontal line represents average value for 
$\Omega_0=6$. Note that the results for $\Omega_0=4$ and $8$ are multiplied by factor of $5$ for 
clarity. Inset: $M^z(t)$ induced by $\Omega_0=\Omega_R=6$ (as in the main panel) for $t$ up to 
$t=250$. (b) Heat map of average net magnetization, $\overline{M^z}-M^z(t=0)$, as a function of 
magnetic field $h$ and frequency $\Omega_0$, calculated for $L=10\,,A=0.1$. $\omega_B$ line (red 
color in the heat map) is obtained with $\Omega>0$ in \eqref{s1t}, $\omega_{A,G}$ (blue color) with 
$\Omega<0$. Solid and dashed lines represent the $\omega_{A,B,G}$ ESR resonance lines and their 
continuation into the gapless regime. Vertical solid line represents the critical field $h_1$.}
\label{fig1}
\end{figure}

\begin{figure}[!ht]
\includegraphics[width=0.9\columnwidth]{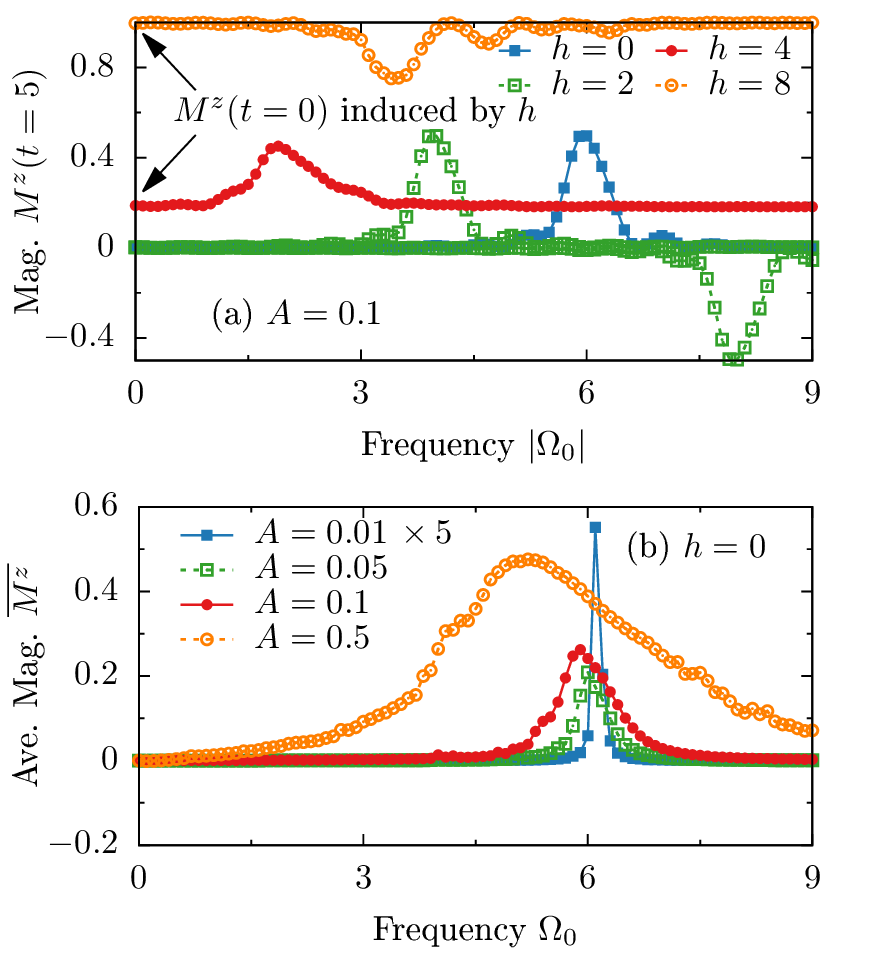}
\caption{(Color online) (a) Frequency $\Omega_0$ dependence of the magnetization $M^z$ at time $t=5$ and 
$L=11$. Results for $h=0,2,4$ are calculated with $\Omega_0>0$ (positive magnetization) and for 
$h=2,8$ with $\Omega_0<0$ (negative magnetization). Note that for $h>h_1$ (presented for $h=4,8$) 
the ground state has net magnetization already at $t=0$. (b) Frequency dependence of average 
magnetization $\overline{M^z}$ for $h=0$, $L=11$ and various amplitudes $A=0.01,0.05,0.1,0.5$. 
Results for $A=0.01$ are multiplied by factor of $5$ for clarity.}
\label{fig2}
\end{figure}

Fig.~\ref{fig1}(a) depicts a typical example of the time dependence of $M^z$ as a function of time for a 
system with $h=0$ and constant $\Omega$. Several conclusions can be drawn directly from the obtained 
results: (i) It is evident that the $M^z$ induced by $\Omega_0=\Omega_R$ is dominating above other 
frequencies. (ii) The beating frequency presented in the inset of Fig.~\ref{fig1}(a) is attributed 
to finite size effects. (iii) The value of $\Omega_R=6$ for $h=0$ is consistent with the lowest 
transition lines of ESR spectrum. In fact, in the gapless regime $h<h_1$, the ESR lines can be 
calculated by a $1/D$ expansion \cite{Papanicolaou1997,Psaroudaki2012}, i.e.,
\begin{equation}
\omega_A=D+2J+h\,,\quad \omega_B=D+2J-h\,.\nonumber
\end{equation}
Such lines correspond to transitions from the GS to states with $\Delta S^z=\pm1$.

In Fig.~\ref{fig1}(b) we present a heat map of the average (over time span $\delta t<t<100$) net 
magnetization, $\overline{M^z}-M^z(t=0)$, as a function of magnetic field $h$ and frequency 
$\Omega_0$. Our results perfectly reproduce both ESR predictions, e.g. see Fig.~6 of 
Ref.~\cite{Psaroudaki2012}. In the considered field $h$ region we also see continuation of the 
$\omega_G=D+h$ line - transitions from a magnon to a single--ion bound state. Other resonance lines 
can also be captured, e.g. transitions from the fully ordered ferromagnetic state in the $h>h_2$ 
region, can be resolved by looking for $\Omega_R$ of negative magnetization. In Fig.~\ref{fig2}(a) 
we present $M^z(t=5)$ as a function of frequency $\Omega_0$. The maximum value of magnetization for 
given $h$ and $\Omega>0$, or $\Omega<0$, is consistent with the ESR predictions.

Although we chose the GS as the starting point of the time evolution this is not a zero temperature 
($T=0$) result. Within LR theory one would expect for $T=0$ rather sharp transition lines \cite{Psaroudaki2013}. 
It is clear from Fig.~\ref{fig1}(b) that our resonance lines are not $\delta$--peaks, with nonzero intensity for 
all considered transitions $\omega_{A,B,G}$. Also, in Fig.~\ref{fig2}(b) we present the dependence 
of $h=0$ average magnetization $\overline{M^z}$ on frequency $\Omega_0$ for various amplitudes 
$A=0.01,0.05,0.1,0.5$ in \eqref{s1t}. Within LR such a broadening of the line could be interpreted 
as the increase of an effective temperature. 

Next, in order to induce a macroscopic magnetization in a controlled way we study the application of a 
chirped pulse, $\Omega=\Omega(t)$. Although the time dependence of $\Omega$ can be complicated and 
its functional form dependent on the experimental setup, the main features should be captured by the 
simple form \begin{equation} \Omega(t)=\Omega_I-\nu t\,, \label{omte} \end{equation} where $\Omega_I$
is the initial ($t=0$) frequency and $\nu$ is the chirp, i.e., the ``speed'' of frequency change. 
Within such a notation $\Omega_I=\Omega_R$ and $\nu=0$ corresponds to a time independent $\Omega$ at 
the resonance frequency. In the following, we will consider only the $h=0$ case, i.e, $\Omega_R=6$, 
as we would like to study the magnetization induced only by light.

\begin{figure}[!ht]
\includegraphics[width=0.9\columnwidth]{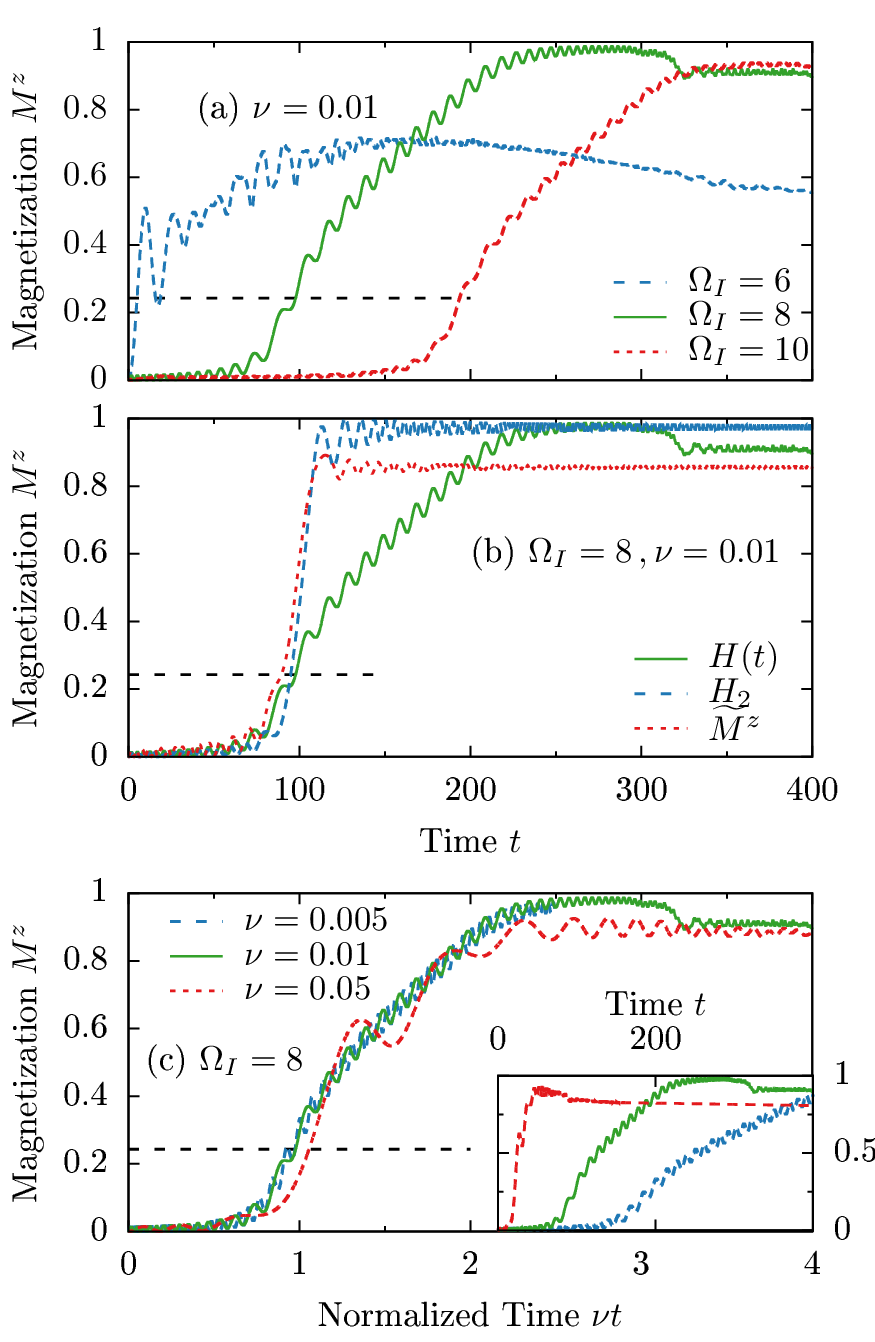}
\caption{(Color online) Time dependence of the magnetization calculated for $L=11$, $h=0$, $A=0.1$. 
(a) Magnetization as a function of time calculated for various initial frequencies $\Omega_I=6,8,10$ and 
$\nu=0.01$. The horizontal line represents the average value of magnetization for $\nu=0$ 
and $\Omega_I=\Omega_R=6$ (resonance frequency). 
(b) Comparison of the magnetization as calculated with Eq.~\eqref{magev} for, (i) the
full Hamiltonian \eqref{s1t} with $\Omega_I=8\,, \nu=0.01$, (ii) the $2$--level model 
\eqref{2level} and the perturbative solution $\widetilde{M}^z$ - Eq.~\eqref{tms} with $\Delta=2$, $\nu=0.01$. 
(c) Magnetization as a function of normalized time $\nu t$ for $\nu=0.005,0.01,0.05$, 
initial frequency $\Omega_I=8$ and $A=0.1$. Inset: the same results as a function of time $t$.}
\label{fig3}
\end{figure}

The qualitative dependence of the magnetization on the amplitude $A$ and chirping $\nu$ can be 
understood within a $2$--level model,
\begin{equation}
{H_2}=0|0\rangle\langle 0|+D|1\rangle\langle 1|+\sqrt{2}A\left(\mathrm{e}^{-\dot{\iota}\Omega t}|1\rangle\langle 0|+\mathrm{h.c.}\right)\,, \label{2level}
\end{equation} 
where $|0\rangle$ ($|1\rangle$) correspond to the $S_i^z=0(1)$ states of the term $D(S^z_i)^2$, relevant to 
the $J/D\to 0$ limit of \eqref{s10}. Note that the resonance frequency of \eqref{2level} is simply
$\Omega_R=D$. 
Within this model, a perturbative expression, $\alpha=A/\sqrt{\nu}\to0$, of the time dependence 
of the magnetization can be given as, 
\begin{eqnarray}
\widetilde{M}^z(t)&=&\frac{|W(t)|^2}{1+|W(t)|^2}\,,\nonumber\\
W(t)&=&\sqrt{2}A\int\limits_0^t\,\mathrm{d}t^{\prime}\,\mathrm{e}^{-\dot{\iota}(\Delta-\nu t^{\prime})t^{\prime}}\nonumber\\
&=&\sqrt{2}\alpha\,\mathrm{e}^{-\dot{\iota}\Delta^2/4\nu}\int\limits_0^{t\sqrt{\nu}}\,\mathrm{d}\tau\,\mathrm{e}^{+\dot{\iota}(\tau-\frac{\Delta}{2\sqrt{\nu}})^2}\,,
\label{tms}
\end{eqnarray}
where $\Delta=\Omega_I-\Omega_R$. It is obvious from the above equation that $\widetilde{M}^z(t)$ 
depends only on $\alpha$ and the detuning $\Delta$.

In Fig.~\ref{fig3}(a) we present the magnetization dependence on the initial frequency $\Omega_I$ at 
fixed $\nu$. In panel (b) we present numerical results obtained from the full Hamiltonain 
\eqref{s1t}, the  $2$--level model \eqref{2level} with corresponding detuning, together with the 
perturbative solution Eq.~\eqref{tms}, that captures the main features of the magnetization profile. 
Note that the results are indistinguishable till the saddle point of Eq.~\eqref{tms}, i.e., at 
$t_s=\Delta/2\nu$. From the results presented in Fig.~\ref{fig3}(b) it is obvious that the main 
effect of the exchange coupling $J$ is in the dynamics of the magnetization at times $t$ beyond 
$t_s$. It is also interesting to note that the magnetization induced by a constant frequency light 
$\Omega=\Omega_R$, as indicated by a dashed line in Fig.~\ref{fig3}, is reached at the saddle point 
time $t_s$. We observe such a behaviour for all $\Omega_I>\Omega_R$. 

In Fig.~\ref{fig3}(c) we show the time dependence of the induced magnetization for different 
chirping speeds $\nu$. We observe that, (i) in the scaled time $\nu t$ the curves are practically 
identical with the crossing of the mean value at $\nu=0$ and $\Omega_I=\Omega_R$ at 
$\nu t \sim \nu t_s=1$, reaching maximum at $\nu t\sim 2$; (ii) the magnetization at long times is 
weakly dependent on time, a remarkable result considering that we are dealing with a full many-body 
problem, where a decay could be expected; (iii) it is clear that, as the total magnetization 
$S^z=\sum_i S_i^z$ commutes with the Hamiltonian, after switching off the light at a certain time, 
$M^z$ remains constant at its instantaneous value. This allows for a tight control of the value of 
the induced magnetization in the system. Further simulations for different $\Omega_I > \Omega_R$ 
confirm this picture; crossing the resonance frequency by chirping the light frequency induces a 
stable macroscopic magnetization in the system. Additionally, it is clear from the solution of the 
$2$--level model that inverting $\Delta'=-\Delta$ and $\nu'=-\nu$ produces identical evolution of 
the magnetization.

Considering the amplitude and chirping speed dependence of the long time asymptotic magnetization 
achieved, first of all we observe that the $2$--level model can be mapped in a rotating frame to a 
Landau--Zener type tunneling problem,
\begin{equation}
\widetilde{H}_2=\widetilde \Delta|\widetilde 1\rangle\langle\widetilde 1|-\widetilde \Delta|\widetilde 0\rangle\langle\widetilde 0|
+\sqrt{2}A\left(|\widetilde 1\rangle\langle\widetilde 0|+\mathrm{h.c.}\right)\,,
\end{equation}
where $\widetilde \Delta=\Delta/2-\nu t$. In the Landau--Zener problem $\Delta=0$ and the time 
evolution is from $t=-\infty$ to $t=+\infty$, while in the situation we are considering the time 
evolution starts at $t=0$ and from a finite frequency shift $\Delta$. For $\Delta/\nu \gg 1$ the 
asymptotic $\widetilde{M}^z(\infty)$ coincides with the probability of occupation of level
$|1\rangle$ given by the Landau--Zener expression $1-\exp(-\pi\alpha^2)$. 

\begin{figure}[!ht]
\includegraphics[width=0.9\columnwidth]{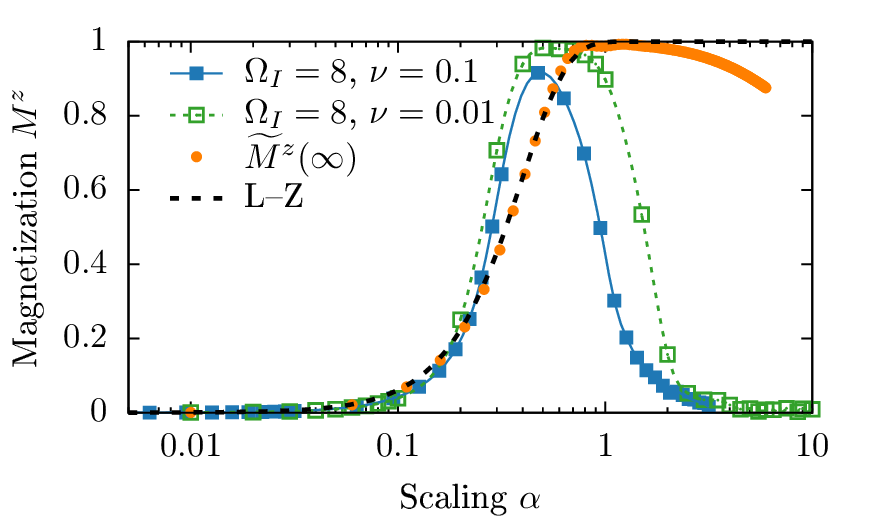}
\caption{(Color online) Scaling parameter $\alpha=A/\sqrt{\nu}$ dependence of the magnetization as 
calculated for $L=11$, $\nu=0.1$ (full squares) and $\nu=0.01$ (open squares). The snapshot of 
magnetization of the full model (squares) is taken at $\nu t=2$, i.e, $t=20$ for $\nu=0.1$ and
$t=200$ for $\nu=0.01$. Circles represent the magnetization $\widetilde{M}^z(\infty)$ dependence on 
$\alpha$ in the $2$--level model \eqref{2level} and the black dashed line depicts the Landau--Zener expression.}
\label{fig4}
\end{figure}

In Fig.~\ref{fig4} we present a comparison of the long time magnetization ($\nu t=2$) in the full model 
\eqref{s1t}, the perturbative prediction Eq.~\eqref{tms} and the Landau–-Zener expression. Note 
that although $\widetilde{M}^z$ is a perturbative solution ($\alpha\to 0$) and the detailed dynamics 
beyond $t_s$ is not captured correctly (see Fig.~\ref{fig3}(b)), the overall agreement of the 
asymptotic magnetization is qualitatively described till $\alpha\sim1$ \cite{3435}. 


Finally, turning to the experimental realization, for the DTN compound ($J=2.2\,\text{K}$, 
$D=8.9\, \text{K}$) the resonance frequency is $\Omega_R\approx 300\,\text{GHz}$. Light of magnetic 
field intensity $\approx0.3\,\text{Tesla}$ corresponding to an electric field of 
$\approx 1\,\text{MV/cm}$ and a chirping speed $\nu\approx 0.1$ will induce a controlled macroscopic 
magnetization within $\approx 1\,\text{psec}$ \cite{pvl}. In a realistic experimental situation, 
several issues arise, (i) in terahertz spectroscopy the light is in the form of a pulse of duration 
$\approx 1\,\text{psec}$, (ii) the effect of electric field should be estimated, (iii) experiments 
are at a finite temperature, (iv) there is spin--lattice relaxation which could be detrimental to 
the process of inducing a macroscopic magnetization. However, it is known that in several quantum 
magnets \cite{Montagnese2013} the relaxation time is surprisingly long. Preliminary finite 
temperature simulations and considerations are encouraging in rendering the proposed experiment 
feasible. We should also note that the large variety of quantum magnets, allow for a tailoring of 
the experiments in terms of light frequency, relaxation time etc. 

In summary, we have studied an efficient protocol which induces magnetization without external 
magnetic field applied to the system. Results for circularly polarized light pulse at constant 
frequency are explained with the help of resonance lines of ESR transitions at finite temperature. 
We have also 
presented comprehensive results on the dependence of the magnetization on a chirped pulse. The 
latter, experimentally relevant, protocol can be qualitatively and even for some time scales 
quantitatively described with the help of a $2$--level model. Also, it was shown
\cite{Psaroudaki2013} that \eqref{s10} can be mapped to an effective $S=1/2$ AHM with exchange 
anisotropy $<1$. Our $2$--level predictions for this model will be even more accurate since the 
mapping favors the large--$D$ 
limit. 

\acknowledgements{This work was supported by the European Union program FP7-REGPOT-2012-2013-1 under 
grant agreement n. 316165 and by the European Union (European Social Fund, ESF), Greek national 
funds through the Operational Program ``Education and Lifelong Learning'' of the NSRF under ``
Funding of proposals that have received a positive evaluation in the 3rd and 4th call of ERC Grant 
Schemes''. We acknowledge helpful and inspiring discussions with R.~Steinigeweg, W.~Brenig,
P.~Prelov\v{s}ek, Z.~Lenar\v{c}i\v{c}, S.~Miyashita, P.~van~Loosdrecht, M.~Montagnese, B. B\"uchner, 
C.~Hess, V. Kataev, S. Takayoshi and T. Oka.}



\begin{thebibliography}{99}
\bibitem{3435} We should note that in Ref.~\onlinecite{Takayoshi1} a similarly driven system 
with weak easy-axis anisotropy and no chirping was considered resulting in a rather small induced magnetization, 
while  in Ref.~\onlinecite{Takayoshi2}, a description in terms of the Landau--Zener tunneling problem 
was presented in systems with $\Omega(t)=\nu t$, $\nu\sim 10^{-4}$ and 
a small symmetry breaking anisotropy term giving $\Omega_R={\cal O}(0.1)$.
\end{thebibliography}
\end{document}